# Study of the photocycle of the metastable states of SRII and their mutants with the use of light activated NMR spectroscopy

Aikaterini Rousaki

*rousaki@umich.edu*

**ABSTRACT**: *Sensory rhodopsin II (SRII) is a seven-helix protein that belongs to the rhodopsin protein family. Light-induced conformational changes govern SRII's function. These changes are related to the photo cycle of the protein that is comprised of various metastable states. After the completion of this cycle the protein returns to its ground state. Mutational studies of key residues will reveal the mechanism that underlies the function of the protein. The result will allow us to determine key structures at various PHs involved in the photo cycle of SRII and to understand the protein's function and mechanism.*

## 1. Subject and aims.

The seven-helix rhodopsin proteins, is a family of membrane proteins that contain retinal as a chromophore and sensory rhodopsin II (SRII) is a member of this class. The members of this family that function as proton pumps and as photoreceptors have been studied. Examples of such proteins are proteorhodopsin[1], bacteriorhodopsin[2] and sensory SRII (also known as phoborhodopsin[3]).

SRII is a protein that has been studied by various laboratories. A short literature- review with the most important publications in the field will be described. The structure of the protein was solved by Pr. Nietlispach.

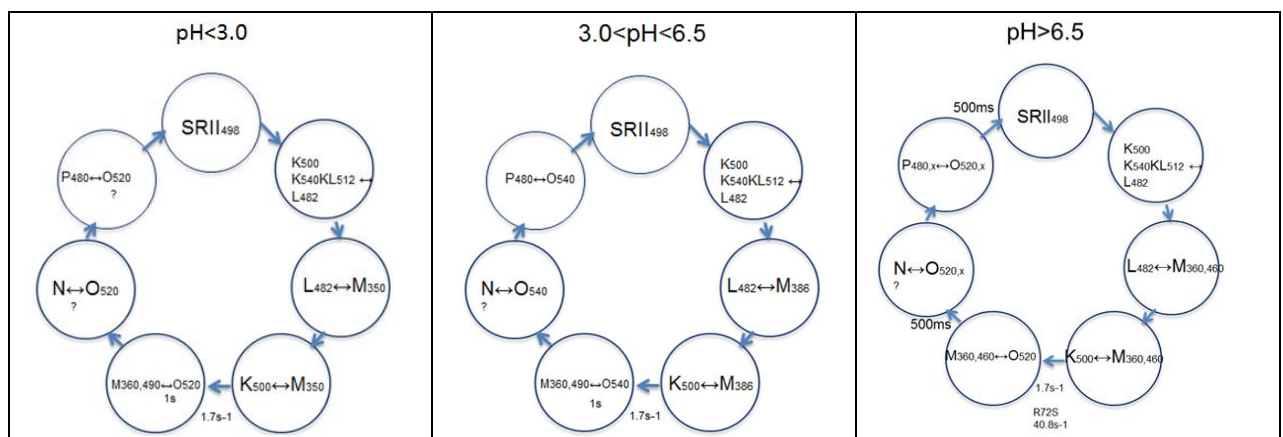

**Figure 1:  SRII photocycle at various PH values.**

All rhodopsins are light-activated proteins. In their photo cycle, short-lived metastable intermediate states (K, L, M, N, O, P) are populated (Figure 1). We choose to picture the SRII photo cycle in these 3 PH ranges because of the following reasoning.  The bibliography[5] indicates that the M intermediate below PH ~6.5 shows an one-phase decay, while above PH ~6.5 it shows a two-phase decay and that the pKa value of the Schiff base counter-ion is 3.0. Therefore the photo cycle is described with three different representations (Figure 1).

| Table 1: The properties of the various states of SRII. | |
|---|---|
| Ground | Light illumination→All trans retinal →all cis retinal. The Schiff base is protonated |
| K | Formation of K state upon light illumination |
| L | Transition of Schiff base proton to Asp75→transition from K to L to $M_1$ |
| M ($M_1$and $M_2$) | When PH>6.5 the $M_1$ state triggers the formation of the $M_2$ state and the formation of the N state follows.  The M and the N states are related to an outward tilting of helix F. |
| N | Ambiguities regarding its existence occur. Multi-exponential analysis suggested its existence – FTIR suggested that it cannot be formed. |
| O | The intermediate O and P are long lived |
| P | The decays times of O and P are in the range of seconds. |

Upon light illumination the SRII all-trans retinal turns to 13-cis and that is the trigger of the SRII photo cycle. The Schiff base is protonated and the K state is formed.  The K state is the first metastable state of the photo cycle and its absorbance maximum is found at 500nm. During the transition of the Schiff base proton to Asp75 the K to L to $M_1$ transition takes place[6,7] .The $M_1$ state triggers the formation of the $M_2$ state (PH>6.5) and the formation of the N state follows. The M and the N states are related to structural changes of helix F and more specifically with an outward tilting of helix F. Regarding the existence of the N state of SRII there are conflicting results in the literature. Multi exponential global analysis suggested that the N intermediate exists[1] while FTIR (Fourier Transform Infrared spectroscopy) suggested that the N intermediate cannot be found [2,3].The intermediates O and P are long lived and have decay times in the range of seconds. After that the system returns back to its ground state.

The lifetime of these states varies among the rhodopsin proteins, their mutants and the conditions of their study.

## 2. Methodology

### 2.A) Laser illumination

In-situ laser illumination and time-resolved NMR experiments will be applied to determine experimental conditions to reveal all the metastable states. In situ laser illumination and subsequent NMR detection will be utilized to trigger the photo cycle of SRII.

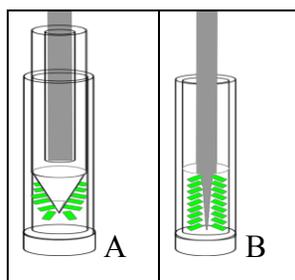

**Figure 3: Optimizing the radiation distribution in the NMR sample. Panel A: 300ul heating-good illumination Panel B:500ul heating-good illumination-big sample volume**

In order to deliver light uniformly into the tube from above with no modifications of the NMR probe, various techniques have been tested (Figure 3). We used the one shown in Figure 3 panel B because it provides good illumination of the sample. After the illumination of the sample the NMR spectrum of the metastable states of SRII are recorded. The experimental pulse sequence that has been implemented combines radio frequency and laser pulses (Figure 4). A long delay time (10 seconds) in the beginning ensures that the system has returned to the ground state. The laser illumination time remains the same for all the experiments and the only parameter that changes is the delay time after illumination. That enabled us to capture three possible intermediate states for the 150, 300 and 600ms mixing times respectively.

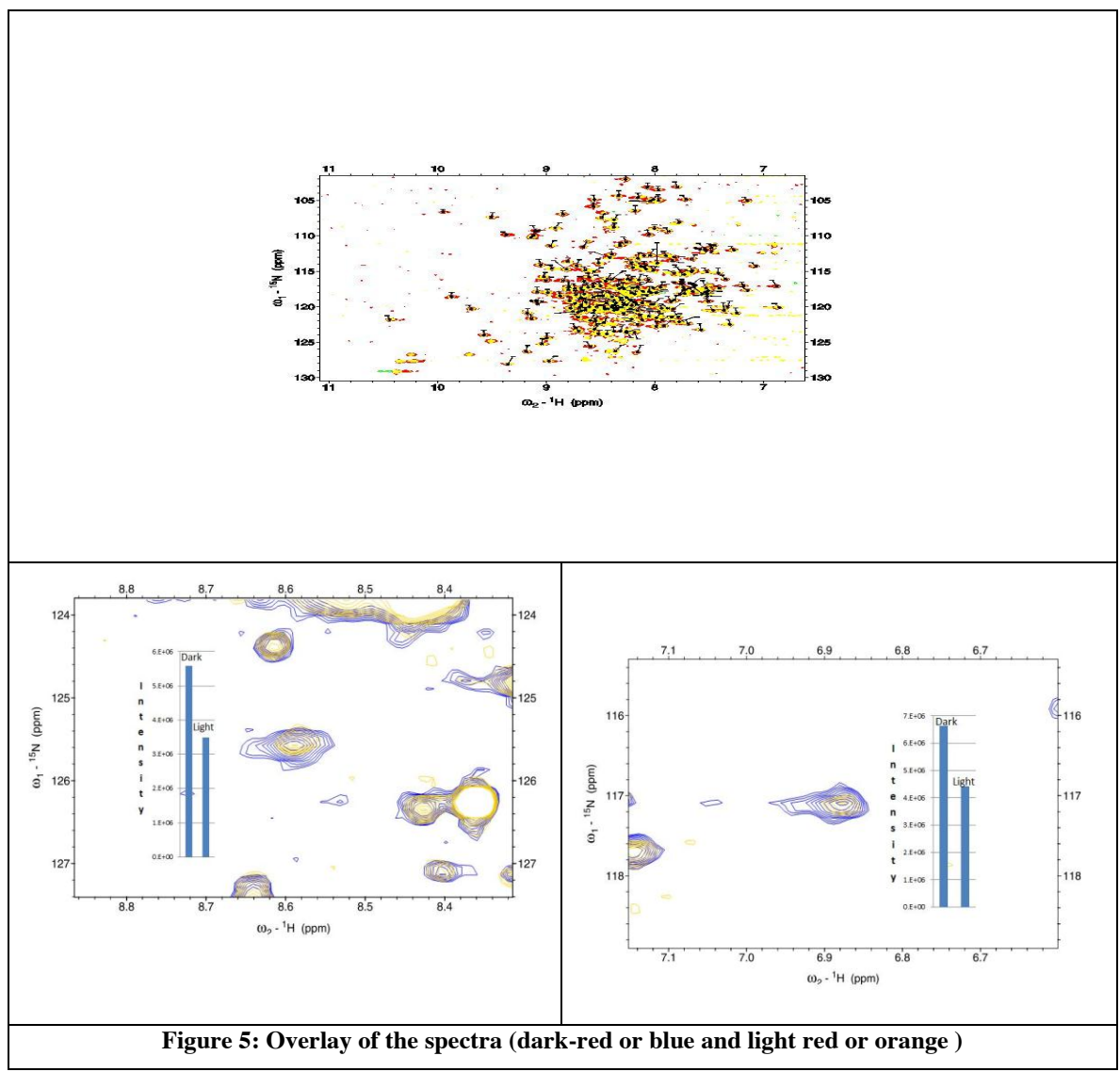

**Figure 5: Overlay of the spectra (dark-red or blue and light red or orange )**

We can discriminate three different metastable states for the three mixing times without being able to define which state each of them represents without the use of further experiments and additional spectroscopic techniques. Further laser NMR experiments that will span the whole timeframe of the photocycle (0 to 3 seconds) will correlate the mixing times with the identities of the metastable states of the photocycle. IR and RAMAN experiments will also have to be implemented in order to achieve that. Also patterns of parallel amino acids in the parallel helices seemed to have the same behavior when the intensity ratio was considered (Figure 7). We suggest varying the delay time in steps of 100ms from the value 0ms till the value 3s, so that we can capture the whole range of the photo cycle and ensure that we obtain all the metastable states.

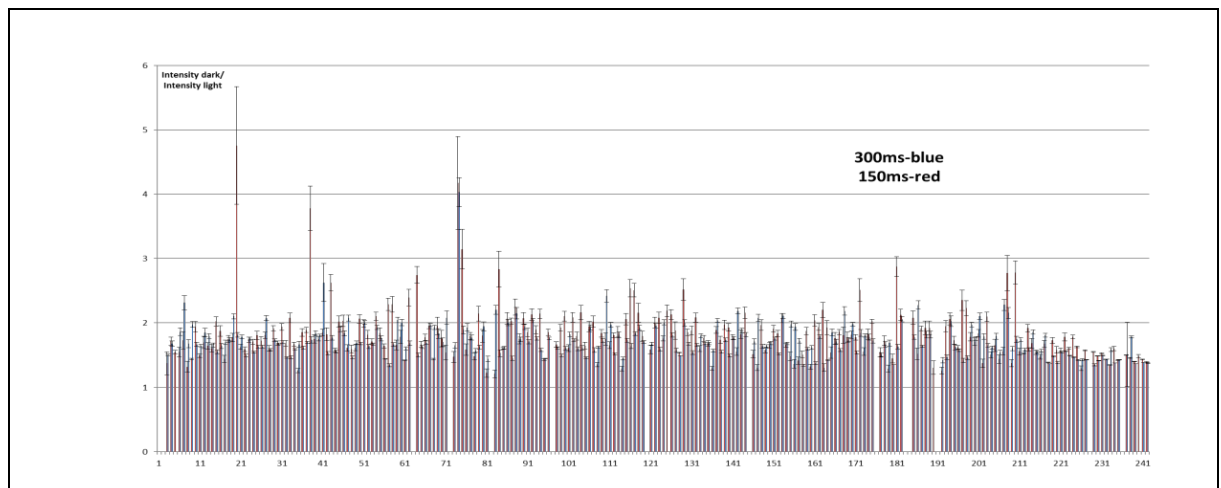

**Figure 6: Overlay of the spectra for $\tau_M$=150 and 300ms that are shown in red and blue respectively. Changes are observed for the residues in direct contact with the retinal: 83(major),109(significant) and 171 (small), which makes us hypothesize that for $\tau_M$ values less than 150ms residue 171 that serves as the trigger of the photo cycle[11] will have major changes present.**

The above data show intensity differences in specific amino acids upon laser illumination that vary with the delay after the laser pulse. The chemical shifts were insignificant and the above data suggest possible relaxation and cross correlation effects taking place. Those effects are possibly dependent on the orientation of the side chains of the amino acids and the backbone of the amino acids comprising the 7 helices of the transmembrane protein.

## 2.B) Suggested experiments in order to better understand the mechanism of the photocycle.

In order to validate the fact that the protein indeed has returned to the dark state after 10 seconds we need to allow 10 seconds as delay time after the laser pulse and expect to get the dark state. In addition in order to determine the time that the photocycle lasts which is ambiguous in bibliography we need to vary the delay time after the laser pulse in steps of 100ms until the system returns to the dark state. When the system returns back to the ground state we can calculate the delay plus the readout TROSY time and determine the time that the photocycle lasts.

### 3. Mutations

Taking into consideration the experimental results (Figure 5, 6 and 8) we were able to suggest some mutations that will enable us to study further the mechanisms of SRII. These mutations can be found in Table 2.

| | A | B | C | D | E | F | HelixG |
|---|---|---|---|---|---|---|---|
| A | | | | | | | E14/V206 (0.172)orL202 (0.25), T6/V192 (-0.101) |
| B | Y37/G26, (0.068), G45/T19 ( -0.18) L56/W9 (0.296),aa/F8 | | | | | | Y37/A212( 0.078), L40/G209( 0.001) |
| C | | F86/L40( 0.059) ,L82/G42 (0.001), D75 /146( -0.006), I74/V49 (0.043), | | | I83/T167 (−0.028) | | F86/A212 (0.06), I83/aa |
| D | | | F98/L93 (0.059), V101/L93 (0.023), M109/aa, L110/I74 (0.078), A114/V70( -0.071) | | | L93/Y160 (0.214),V101/T167-(0.101) | |
| E | | | | E147/L93 (0.25),Y139/V101( 0.042),A125/G115( - 0.322) orA114(-0.269), R123/A116 (0.578)orG115 (-0.496)orA114( 0.578), E122/A114 (0.239) | | | |
| F | | | | | Y160/T146 (0.022), T167/L141 (0.007), W171/aa, Y174/aa | | |
| G | | | | | | T218/E155 −(0.18), aa/R162,L200/Y174 (0.214),L196/aa, V192/L180 (0.023) | |
| | Helix A | Helix B | Helix C | Helix D | Helix E | Helix F | Helix G |
| retinal | | | I83 | | A131 | W171,Y174 | K205, D201 |
| β sheet | | | V161/V70 | | | | |

**Table 2: Suggested mutations based on the intensity changes between the states with delay times 150 and 300ms. aa stands for any amino acid close to the suggested partner.**

These mutations have the potential to interfere with the mechanism and function of SRII, considering the fact that they represent the amino acids that are affected upon transition of the system from the state represented with mixing time 150ms to the state represented with mixing time 300ms. Some mutations might be more potent than others. A recent paper[13] describes how favorable the amino acid interactions in helical membrane proteins are, considering how often various amino acid interactions are found in helical membrane proteins. We wish to note here that the favorability of amino acid interactions in membrane proteins and proteins in solution differ significantly. The amino acid favorability is described by a scoring matrix that contains values with positive and negative signs. This matrix has been obtained from[13] and can be seen in Table 3. The scoring matrix values were obtained with the use of the information theory. According to that a training set of 84 helical membrane proteins with known native state structures was taken. The different types of pair-wise contacts were counted. These numbers were compared to a random reference set. If the number of contacts exceeded that in the reference set, the interaction was defined as attractive and the strength of the interaction is related to the degree of excess. In analogy a repulsive interaction was defined.

When the interaction is favorable the value is low e.g. a Lys-Lys interaction has a value of -1.015 and represents a very favorable interaction in α-helical membrane proteins. In analogy when the matrix value is high the interaction is unfavorable. These values for the amino acid interactions of interest are given in parenthesis in Table 2. It is noteworthy to mention that the Lys205 is close to Thr167 (scoring value: 0.5), Ile83 (scoring value: 0.41) and Ile43 (scoring value 0.41). These interactions are highly unfavorable but the functional importance of Lys205 is unquestionable since it is protonated by the retinal and triggers the cycle[14]. Also note that Thr167and Ile83, change their intensities significantly upon transition from the 150ms to the 300ms state (Figure 8 and 9). So our difference in intensity analysis is further confirmed.

Considering the fact that with the use of scoring matrices and residues that show significant intensity changes upon transition from the 150ms to the 300ms state we could identify the Lys205 which is the amino acid that gets protonated by the retinal chromophore, we could use the same approach to identify and other residues that affect significantly the cycle.

In Table 2 the underlined amino acid interactions are unfavorable according to the matrix theory and are all very good targets for mutations. The ones that stand out though are Glu115 of helix D that seems to stabilize the interaction with helix E. If we check the value for the Ala125/Gly115 interaction, which is favorable with a value of -0.322 and transform it to unfavorable by mutating

Gly115 to Ala (0.578), then we will have 3 Alanines in a row: Ala114, Ala115(mutation from Gly115), Ala116 interacting with Ala 125. That will inhibit the interactions of Helices D. A neighboring mutation that could take place is that of Arg123 with Ala116 (0.578), Gly115(0.496) and Ala114 (0.578). All these interactions are strongly unfavorable indicating the significance of residue Arg123. We suggest the mutation of Arg123 to a Glycine in order to make the interactions of Ala/Gly -0.322 and Gly/Gly -0.653 favorable. That will diminish the function of Arg123 and we will be able to understand the significance of that amino acid in the cycle. Also the neighboring interaction of Glu122 with Ala114 (0.239) is unfavorable and a potential target for mutational studies. We could make that interaction more unfavorable by substituting Glu122 with Leu (Leu/Ala 0.25), Ile (Ile/Ala 0.277), His (His/Ala 0.498), Ser (Ser/Ala 0.352) or by substituting Ala114 with Arg (Glu/Arg 0.578). In analogy we could make that interaction favorable by substituting Glu122 with Trp (Trp/Ala -0.456),Gln (Gln/Ala -0.517), Arg (Arg/Ala -0.522), Asn (Asn/Ala 0.573) or by substituting Ala114 with His (Glu/His -0.254) or Gly (Glu/Gly -0.322).

**Table 3: Scoring matrix as obtained from the Supplementary material of** [33]

| L | F | I | M | V | W | C | Y | H | A | T | G | P | R | Q | S | N | E | D | K |
|---|---|---|---|---|---|---|---|---|---|---|---|---|---|---|---|---|---|---|---|
| -0.017 | 0.059 | 0.078 | 0.219 | 0.023 | 0.296 | -0.072 | 0.214 | 0.114 | -0.052 | 0.007 | 0.001 | 0.159 | 0.296 | 0.263 | -0.079 | 0.084 | 0.25 | 0.221 | 0.454 |
| 0.059 | -0.137 | 0.045 | -0.072 | 0.104 | 0.151 | 0.029 | 0.077 | 0.19 | 0.06 | 0.058 | -0.103 | 0.009 | 0.13 | 0.096 | -0.006 | 0.169 | 0.134 | 0.276 | 0.077 |
| 0.078 | 0.045 | -0.087 | -0.003 | 0.043 | 0.341 | -0.057 | 0.178 | 0.066 | 0.061 | -0.028 | -0.064 | -0.06 | 0.3 | 0.253 | -0.032 | 0.11 | 0.277 | -0.006 | 0.41 |
| 0.219 | -0.072 | -0.003 | -0.168 | 0.056 | 0.296 | -0.102 | 0.231 | 0.037 | -0.057 | -0.076 | -0.215 | 0.17 | 0.094 | -0.193 | -0.151 | 0.383 | 0.091 | 0.18 | 0.178 |
| 0.023 | 0.104 | 0.043 | 0.056 | 0.02 | 0.073 | -0.002 | 0.042 | 0.055 | -0.071 | -0.101 | -0.174 | 0.114 | 0.46 | -0.076 | 0.006 | 0.026 | 0.172 | 0.246 | 0.371 |
| 0.296 | 0.151 | 0.341 | 0.296 | 0.073 | -0.204 | -0.359 | -0.004 | -0.302 | 0.1 | 0.469 | 0.045 | 0.274 | -0.157 | 0.41 | 0.152 | 0.204 | -0.122 | 0.176 | -0.122 |
| -0.072 | 0.029 | -0.057 | -0.102 | -0.002 | -0.359 | 0.557 | 0.193 | 0.222 | -0.133 | -0.118 | -0.03 | 0.421 | 0.38 | 0.051 | -0.291 | -0.188 | -0.456 | -0.456 | 1.467 |
| 0.214 | 0.077 | 0.178 | 0.231 | 0.042 | -0.004 | 0.193 | -0.222 | 0.38 | 0.078 | 0.022 | 0.068 | -0.018 | 0.202 | 0.037 | 0.037 | -0.086 | 0.035 | 0.267 | -0.03 |
| 0.114 | 0.19 | 0.066 | 0.037 | 0.055 | -0.302 | 0.222 | 0.38 | -0.709 | -0.254 | 0.117 | -0.161 | 0.838 | -0.21 | 0.43 | -0.103 | -0.16 | 0.498 | -0.261 | 0.994 |
| -0.052 | 0.06 | 0.061 | -0.057 | -0.071 | 0.1 | -0.133 | 0.078 | -0.254 | 0.578 | 0.509 | 0.496 | -0.006 | -0.382 | 0.609 | 0.243 | -0.333 | -0.522 | -0.332 | -0.447 |
| 0.007 | 0.058 | -0.028 | -0.076 | -0.101 | 0.469 | -0.118 | 0.022 | 0.117 | 0.509 | -0.002 | -0.166 | 0.112 | 0.509 | 0.051 | -0.229 | -0.011 | -0.18 | -0.255 | 0.5 |
| 0.001 | -0.103 | -0.064 | -0.215 | -0.174 | 0.045 | -0.03 | 0.068 | -0.161 | 0.496 | -0.166 | -0.653 | 0.014 | 0.496 | -0.344 | -0.183 | 0.181 | 0.169 | 0.482 | 0.131 |
| 0.159 | 0.009 | -0.06 | 0.17 | 0.114 | 0.274 | 0.421 | -0.018 | 0.838 | -0.006 | 0.112 | 0.014 | 0.643 | -0.006 | -0.304 | 0.07 | 0.416 | 0.049 | -0.417 | 0.846 |
| 0.296 | 0.13 | 0.3 | 0.094 | 0.46 | -0.157 | 0.38 | 0.202 | -0.21 | -0.382 | 0.509 | 0.496 | -0.006 | 0.609 | 0.609 | 0.243 | -0.333 | -0.522 | -0.332 | -0.447 |
| 0.263 | 0.096 | 0.253 | -0.193 | -0.076 | 0.41 | 0.051 | 0.037 | 0.43 | 0.609 | 0.051 | -0.344 | -0.304 | 0.609 | -0.361 | -0.046 | 0.153 | -0.517 | -0.522 | 0.304 |
| -0.079 | -0.006 | -0.032 | -0.151 | 0.006 | 0.152 | -0.291 | 0.037 | -0.103 | 0.243 | -0.229 | -0.183 | 0.07 | 0.243 | -0.046 | 0.115 | -0.284 | 0.352 | 0.283 | 0.189 |
| 0.084 | 0.169 | 0.11 | 0.383 | 0.026 | 0.204 | -0.188 | -0.086 | -0.16 | -0.333 | -0.011 | 0.181 | 0.416 | -0.333 | 0.153 | -0.284 | -0.45 | -0.573 | -0.192 | 0.204 |
| 0.25 | 0.134 | 0.277 | 0.091 | 0.172 | -0.122 | -0.456 | 0.035 | 0.498 | -0.522 | -0.18 | 0.169 | 0.049 | -0.522 | -0.517 | 0.352 | -0.573 | -0.178 | -0.675 | 0.597 |
| 0.221 | 0.276 | -0.006 | 0.18 | 0.246 | 0.176 | -0.456 | 0.267 | -0.261 | -0.332 | -0.255 | 0.482 | -0.417 | -0.332 | -0.522 | 0.283 | -0.192 | -0.675 | -0.718 | -0.718 |
| 0.454 | 0.077 | 0.41 | 0.178 | 0.371 | -0.122 | 1.467 | -0.03 | 0.994 | -0.447 | 0.5 | 0.131 | 0.846 | -0.447 | 0.304 | 0.189 | 0.204 | 0.597 | -0.718 | -1.015 |

Another possible mutation includes the exact antidiametric site of the Ala125/Gly115 and that is Leu56/Trp9 which is considered as an unfavorable interaction among helices B and A with a value of 0.296. We could either transform that interaction to a more unfavorable one by mutating Trp9 to a Lys (Leu/Lys 0.454) or by mutating Leu56 to an Ile (Ile/Trp 0.341), a Thr

(Thr/Trp 0.469) or a Gln(Gln/Trp 0.41) or make that interaction more favorable by mutating Leu56 to a Cys (Cys/Trp -0.359). We could also target for the Leu93/Tyr160 (0.214) interaction which is also unfavorable and either do it more unfavorable by substituting Leu93 with Lys (Lys/Tyr 0.454) or Arg (Arg/Tyr 0.296) or by substituting Tyr 160 with Asp (Leu/Asp 0.283). In addition we could make that interaction more favorable by substituting Leu93 with Tyr (Tyr/Tyr -0.222). All these interactions can be seen in Figure 9 as magenta spheres.

| | |
|---|---|
| Leu56/Trp9 | Ala125/Gly115andAla114 |
| Glu147/Leu93 | Arg123/Ala114,Gly115,Ala116 |
| Leu93/Tyr160 | Gly122/Ala114 |
| Leu200/Tyr174 | Glu14/Leu202 |

**Figure 9: Potent amino acid interactions that could be mutated and affect the photocycle are shown as magenta spheres while in magenta the amino acid that show significant intensity changes have been depicted.**

We can also target the Leu200/Tyr174 (0.214) unfavorable interaction and either make it more unfavorable by substituting Tyr 174 with Lys (Leu/Lys 0.454) or make it more favorable by substituting Leu200 with Tyr (Tyr/Tyr -0.222). In addition we could target the Glu14/Leu202 unfavorable interaction by either making it more unfavorable or more favorable. In order to make it more unfavorable we would need to substitute Leu202 with His (Glu/His 0.498), Glu (Glu/Glu 0.322) or Lys (Glu/Lys 0.597) or to substitute Glu14 with Lys (Lys/Leu 0.454). In order to make the interaction more favorable we would need to mutate Leu202 to Cys (Glu/Cys -0.456), Arg (Glu/Arg -0.522), Gln (Glu/Gln -0.517), Asn (Glu/Asn -0.573).

The result of these mutations (Table 4) will be the identification of key amino acids in the cycle for the transition of the system from the state with mixing time 150ms to the state with mixing time 300ms. Once more states have been identified we could follow the same procedure and determine more key amino acids for the specific transition of the photocycle that we encounter each time.

| Table 4: Suggested mutations | | |
|---|---|---|
| Amino acid interaction | Mutation(unfavorable) | Mutation (favorable) |
| Gly115/Ala125 | G115A | |
| Arg123/Ala116 | | R123G |
| Glu122/Ala114 | E122L,E122I, E122H, E122S, A114R | E122W, E122Q, E122,R, E122N, A114H, A114G |
| Leu56/Trp9 | W9K, L56I, L56T, L56Q | L56C |
| Leu93/Tyr160 | L93K,L93R, Y160D | L93Y |
| Leu200/Tyr174 | Y174K | L200Y |
| Glu14/Leu202 | L202H, L202E, L202K, E14K | E14C, E14R, E14Q, E14N |

# 4. PH impact

We would like to investigate the effect of PH on the SRII protein photo cycle. As indicated in[4] the protein behaves differently in the three PH ranges (Figure 1) and thus we need to investigate the effect of PH in the PH ranges lower than 3, between 3 and 6.5 and greater than 6.5.  That will indicate at which PH range this protein functions best and how many metastable states are present in each PH range. As seen in Figure 1 the M state for a PH range greater than 6.5 is represented by two states $M_1$ and $M_2$ and the literature[13] implies that it is possible that the $M_1$ and $M_2$ states are also present in the 3 to 6.5 PH range That is a question that needs to be answered.

# References and further reading